\documentclass{PoS}

\title{Progress On Neutrino-Proton Neutral-Current Scattering In MicroBooNE}

\ShortTitle{Progress On Neutrino-Proton Neutral-Current Scattering In MicroBooNE}

\author{\speaker{Stephen Pate}%
       \thanks{This research supported by the US Department of Energy, Office of Science, 
       Medium Energy Nuclear Physics.}\\
      (for the MicroBooNE Collaboration)\\
      Physics Department, New Mexico State University, Las Cruces NM 88003, USA\\
      E-mail: \email{pate@nmsu.edu}}

\abstract{The MicroBooNE Experiment at the Fermi National Accelerator Laboratory, an 89-ton active mass liquid argon 
time projection chamber, affords a unique opportunity to observe low-$Q^2$ neutral-current neutrino-proton 
scattering events. Neutral-current
neutrino-proton scattering at $Q^2 < 1$ GeV$^2$ is dominated by the proton's 
axial form factor, which can be written
as a combination of contributions from the up, down, and strange quarks:
$G_A(Q^2) = \frac{1}{2}[-G_A^u(Q^2)+G_A^d(Q^2)+G_A^s(Q^2)]$.  The contribution 
from up and down quarks has been established in past charged-current measurements.  
The contribution from strange quarks at low $Q^2$ remains unmeasured; this is of 
great interest since the strange quark contribution to the proton spin can be determined from the 
low-$Q^2$ behavior:  $\Delta S = G_A^s(Q^2=0)$.
MicroBooNE began operating in the Booster Neutrino Beam in October 2015. I will present
the status in observing isolated proton tracks in the MicroBooNE detector as a signature for 
neutral-current neutrino-proton events.
The sensitivity of the MicroBooNE experiment for measuring
the strange quark contribution to the proton spin 
will be discussed.}

\FullConference{The 26th International Nuclear Physics Conference\\
		 11-16 September, 2016\\
		 Adelaide, Australia}

\begin{document}

\bibliographystyle{siam}

\section{Motivation}

The cross section for neutrino-proton neutral-current scattering~\cite{Garvey:1992cg}, $\nu p \rightarrow \nu p$, 
is determined by the
standard model behavior of neutrinos and the structure of the proton as expressed in the electric, magnetic, 
and axial form factors of the proton; $G_E^p(Q^2)$, $G_M^p(Q^2)$, and $G_A^p(Q^2)$ respectively.  The electric 
and magnetic form factors are well-determined for momentum transfers $Q^2<1$ GeV$^2$ from electron-nucleon 
elastic scattering data~\cite{PhysRevC.76.035205}.
The contribution to the axial form factor from up and down quarks is also well-determined for $Q^2<1$ GeV$^2$ 
by neutrino-deuteron charged-current reaction data~\cite{Bodek:2007ym}, and 
at $Q^2=0$ GeV$^2$ from neutron decay~\cite{Cabibbo:2003cu}.
The dominant unknown is the strange quark contribution to the axial form factor: $G_A^s(Q^2)$.  
A measurement of this form factor at low $Q^2$ can determine the total strange quark contribution to the 
proton spin, $\Delta S \equiv \Delta s + \Delta\bar{s}$, from an extrapolation to $Q^2=0$: 
$\Delta S = G_A^s(Q^2=0)$~\cite{Pate:2003rk}.

The physics interest in the strange quark contribution to the nucleon spin is long-standing and widespread.  In addition
to being a missing piece of the proton spin puzzle, it is also vital for the interpretation of searches for heavy
dark-matter particles~\cite{Ellis:2008hf}.  Three-dimensional simulations of supernovae~\cite{Melson:2015spa} 
are sensitive
to the value of $\Delta S$, as are atomic parity-violation experiments on hydrogen~\cite{Gasenzer:2011hs}.  Recent 
lattice QCD calculations~\cite{PhysRevD.86.114510,PhysRevD.85.054510} 
give small values (less than 0.003 in magnitude) for $\Delta S$; this 
requires experimental verification.

The first experimental data on $\Delta S$ came from measurements of the {\em inclusive} deep-inelastic
scattering (DIS) of polarized muons from polarized hydrogen, in the EMC experiment in the 1980s~\cite{Aidala:2012mv}, 
and indicated a negative value for $\Delta S$.
This has been confirmed in all subsequent inclusive measurements at SMC, SLAC, HERMES, COMPASS, and JLab.  The
analysis of polarized inclusive DIS data always assumes SU(3) flavor symmetry, combining the extrapolated integral
of the DIS measurements with the triplet and octet axial charges determined from hyperon $\beta$-decay.

Later, it became possible to observe {\em semi-inclusive} deep-inelastic scattering, where the leading hadron 
(a pion or kaon) served to ``tag'' the struck quark.  These measurements by SMC, HERMES, and 
COMPASS~\cite{Aidala:2012mv} have 
consistently implied
that $\Delta S$ is consistent with zero, in contradiction to the inclusive measurements.  The analysis of these data
differs strongly from that of the inclusive data, not using SU(3) flavor symmetry 
but instead relying on an understanding of quark$\rightarrow$hadron fragmentation functions.

This dichotomy between the results of the inclusive and semi-inclusive measurements continues to the present day.  
Global analyses~\cite{PhysRevD.80.034030,Nocera:2014gqa,Leader:2014uua,Hirai2009106,Bluemlein2010205} 
of leptonic DIS and polarized $pp$ collision data show this discrepancy in the determination
of $\Delta S$.

An alternative method to determine $\Delta S$ is available from a measurement of the axial form factor of the proton
in {\em elastic} neutrino-proton scattering.  Cross sections for elastic $\nu p$ scattering (from within carbon nuclei)
exist from the BNL-E734 experiment~\cite{Ahrens:1987xe} and MiniBooNE~\cite{AguilarArevalo:2010cx}, 
but neither of these measurements extends 
below $Q^2=0.45$ GeV$^2$ and therefore cannot be reliably 
extrapolated to $Q^2=0$ for a precise 
determination of $\Delta S$.  An analysis
of currently available data was 
explored in detail in \cite{Pate:2008va} and \cite{Pate:2012wg}.  A measurement of neutrino-proton elastic scattering at
low $Q^2$ will make possible a more precise determination of $\Delta S$.

Such a measurement is a difficult pursuit.  The observable part of the final state in elastic $\nu p \rightarrow \nu p$
is a single isolated proton track.  For $Q^2\approx 0.1$ GeV$^2$, the kinetic energy of the proton will be
approximately 50 MeV.  In a liquid or solid detector, such a proton might travel only a few centimeters.  The 
development
of large-scale liquid argon time projection chambers, however, has made the efficient detection of such events,
with good statistics, a realistic possibility.

Another difficulty arises from the use of a nuclear target, in our case the argon-40 nucleus.  In particular,
the proton in the final state may re-scatter with other nucleons before escaping the nucleus.  To mitigate this and
other effects, we have chosen as our observable the neutral-current to charged-current yield ratio (see Section~\ref{section:impact}),
$$R_{NC/CC} = \frac{N(\nu p \rightarrow\nu p)}{N(\nu n \rightarrow\mu p)}.$$
We will need input from nuclear theory to establish the reliability of this (or any other) approach to the effects
the nucleus will have on the neutral-current and charged-current yields.

\section{The MicroBooNE Experiment}

MicroBooNE~\cite{Karagiorgi:2012cba} 
is an accelerator-based neutrino experiment at Fermi National Accelerator Laboratory, 
centered around a liquid argon time projection chamber (LAr TPC) with an active mass of 89 tons of liquid argon 
in a field cage of dimensions (10 m)$\times$(2.3 m)$\times$(2.6 m).  
The TPC was placed into the experimental hall in the summer of 2014, 
installed and commissioned, observed first cosmic rays in the summer of 2015, and began to take data with the Booster 
Neutrino Beam (BNB) on October 15, 2015.  Since that time, in-beam data has been taken using 
$3.4\times 10^{20}$ protons-on-target.

Analysis of the TPC wire data produces images of charge vs. time in the TPC volume, revealing tracks from 
charged-particles that transited the active volume.  With three wire planes, there are three images of each event, 
making possible a three-dimensional reconstruction of the ionization tracks.  In addition to the charged-track images,
a light collection system (composed of 32 8-inch photomultiplier tubes) records flashes of light from the scintillation
of the liquid argon.
There will be millions of these images and associated light flashes recorded over 
the course of the experiment, so traditional ``visual scanning'' of such data is out of the question.  
Automated software must sort through the events; for large-scale LAr TPC data this will be the first time such 
automated sorting has been attempted.

For finding the isolated proton tracks of interest here, Katherine Woodruff of NMSU has led an effort to utilize
a ``boosted decision tree'' technique.  A boosted decision tree is a series of if/then/else questions that is tuned
using ``training data.''  In our case, the questions will be directed at a variety of track features; geometry 
(track length and orientation), calorimetry (total charge, total light), and matching between the track location and 
flash location.  The result of the tree, for each track, is the assignment of probabilities that the track is a member
of five different classes:
a proton, a muon, a pion, an electromagnetic shower, or a cosmic-generated track.

Using a full simulation of the detector, including all detector effects (true geometry, realistic noise, 
missing wires, etc.), we can train the tree to optimize the track classification.  
Then, using a second set of simulated events,
we determine the efficiency of the tree for each class.  
Then we are able to use the boosted decision tree on real in-beam data.

\begin{figure}
\begin{center}
\includegraphics[width=\textwidth]{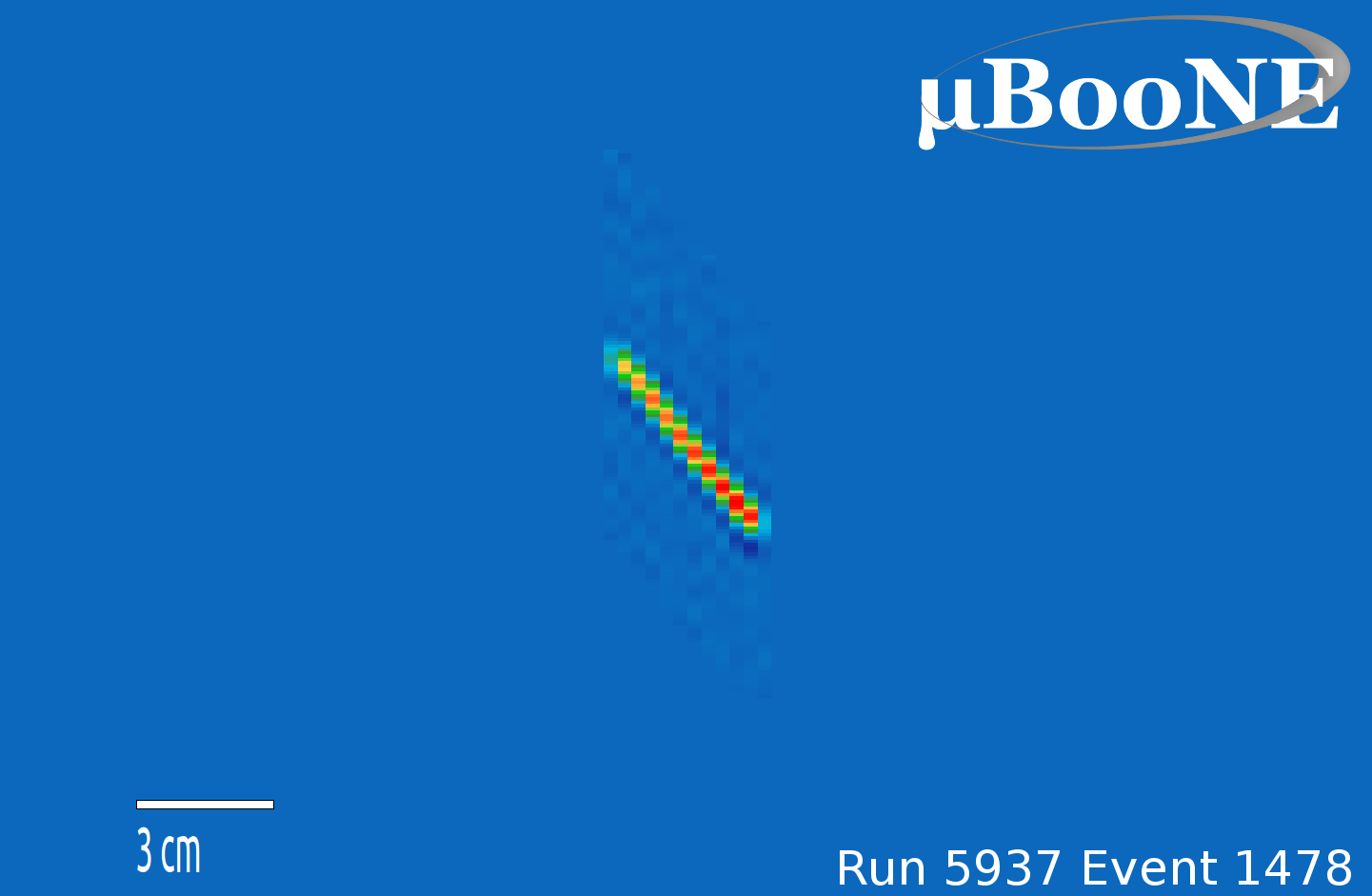}
\caption{A proton candidate track in MicroBooNE data 
selected by a trained boosted decision tree.  The wire spacing is 3mm.
The pixel color is proportional to ionization charge density.}
\label{proton_track}
\end{center}
\end{figure}

Figure~\ref{proton_track} shows a candidate for an isolated proton track that was selected by a boosted decision tree.
The proton probability assigned to this track was 87\%.  This view of the track is from just one wire plane.  Using
all three planes, the reconstructed three-dimensional track length is 5.9 cm, corresponding to a proton kinetic
energy of approximately 82 MeV.  If this were in fact a neutral-current scattering event, the momentum transfer
would be $Q^2\approx 0.15$ GeV$^2$.

\section{Potential Impact}
\label{section:impact}

What will be the impact on our knowledge of $\Delta S$ due to the new neutral-current elastic scattering data from
MicroBooNE?  This question has been studied in some detail and reported already in \cite{Pate:2014kea}; 
a review of that work is given here.  
The central feature of this study is a global fit of elastic electron-nucleon and neutrino-electron scattering data
to extract the strange quark contribution to the electric, magnetic and axial form factors of the proton; 
$G_E^s(Q^2)$, $G_M^s(Q^2)$, and $G_A^s(Q^2)$ respectively.  A simple functional model is used for those three
form factors in this global fit:
$$G_E^s = \rho_s\tau ~~~~~~~~~~ G_M^s=\mu_s ~~~~~~~~~~ G_A^s = \frac{\Delta S + S_A Q^2}{(1+Q^2/\Lambda_A^2)^2}$$
where $\rho_s$ is the strangeness radius, $\tau=Q^2/4M_N^2$, $\mu_s$ is the strangeness contribution to the proton
magnetic moment, and $S_A$ and $\Lambda_A$ are shape parameters that were needed to obtain a good fit to the
existing data.
The result (see Table~\ref{fit_table}) is that $G_E^s(Q^2)$ and $G_M^s(Q^2)$ are
well constrained throughout the range $0<Q^2<1.0$ GeV$^2$ 
by existing data (and are consistent with zero), but $G_A^s(Q^2)$ is unconstrained at low $Q^2$ due to the lack
of neutrino-proton elastic data below $Q^2=0.45$ GeV$^2$.  
\begin{table}[ht]
\centering
\caption{Preliminary results for our 5-parameter fit to the 49 elastic neutrino- and parity-violating
electron-scattering data
points from BNL E734, HAPPEx, SAMPLE, G0, and PVA4.  The first uncertainty arises from the experimental data 
uncertainties; the second uncertainty arises from uncertainties in radiative corrections. Note: these values differ
from reference \cite{Pate:2014kea} due to the inclusion of new data from PVA4 on scattering from deuterium.}
\label{fit_table}
\begin{tabular}{cc}
\hline
Parameter & Fit value \\ \hline
$\rho_s$ & $-0.10 \pm 0.09 \pm 0.03$ \\
$\mu_s$ & $0.056 \pm 0.029 \pm 0.022$ \\
$\Delta S$ & $-0.29 \pm 0.42 \pm 0.19$ \\
$\Lambda_A$ & $1.1 \pm 1.0 \pm 1.1$ \\
$S_A$ & $0.4 \pm 0.5 \pm 0.2$ \\ \hline
\end{tabular}
\end{table}

Taking that fit as a starting point, a 
simulation\footnote{Thanks to B. Fleming, J. Spitz, and V. Papavassiliou for providing this simulation.}
was performed to generate a set of neutral current (NC) $\nu p \rightarrow\nu p$ and charged-current (CC)
$\nu n \rightarrow \mu p$ 
events corresponding to a sample of MicroBooNE data using 
$2\times 10^{20}$ protons-on-target, that is about 1/3 of the data sample that MicroBooNE plans to collect in 
our initial run.
These simulated events were used to estimate the statistical uncertainty in the NC/CC yield ratio,
$$R_{NC/CC} = \frac{N(\nu p \rightarrow\nu p)}{N(\nu n \rightarrow\mu p)}.$$
This ratio is chosen as our experimental observable not only because 
many experimental uncertainties approximately cancel
in this ratio (e.g. proton detection efficiencies, target mass, neutrino flux), but also because some theoretical
uncertainties (e.g. nuclear final state effects on the proton yield) may cancel as well.  This sample does not
represent the full statistics we expect in the experiment, but on the other hand it does not include
the effect of reconstruction efficiencies and backgrounds from neutrons and cosmic rays.  Our intention here
is to qualitatively show the potential impact of new data in the low-$Q^2$ region.

The simulated results for $R_{NC/CC}$ are then treated as actual data and the global fit is repeated.  The change
in the {\rm uncertainties} of the fit parameters are shown in Table~\ref{uboone_table}.  The
fitted mean values for the form factors do not shift because they were the basis of the MicroBooNE event simulation, so 
these are not shown in the table;
however the uncertainties in those form factors are reduced.  The uncertainties in the electric and magnetic
form factors are improved, but not very significantly.  However, the uncertainty in the parameters for the axial form 
factor ($\Delta S$, $S_A$, and $\Lambda_A$) are greatly reduced, by approximately a factor of 10. 
This illustrates the very significant impact that MicroBooNE data on neutral current elastic scattering can 
potentially have
on our knowledge of the strangeness contribution to the axial form factor and to the proton spin.
\begin{table}[ht]
\centering
\caption{Improvement in {\em uncertainties} 
in global fit parameters, when simulated MicroBooNE data are included in the fit.}
\label{uboone_table}
\begin{tabular}{ccc}
\hline
Parameter & Using Existing Data & Including MicroBooNE \\ \hline
$\rho_s$ & $\pm 0.09 \pm 0.03$ & $\pm 0.08 \pm 0.02$ \\
$\mu_s$ & $\pm 0.029 \pm 0.022$ & $\pm 0.023 \pm 0.017$ \\
$\Delta S$ & $\pm 0.42 \pm 0.19$ & $\pm 0.036 \pm 0.003$ \\
$\Lambda_A$ & $\pm 1.0 \pm 1.1$ & $\pm 0.42 \pm 0.03$ \\
$S_A$ & $\pm 0.5 \pm 0.2$ & $\pm 0.05 \pm 0.02$\\ \hline
\end{tabular}
\end{table}

\section{Summary}

The MicroBooNE Experiment has taken data for one year with the Booster Neutrino Beam at Fermilab, the start of a
multi-year data program.  We expect
to have a dataset on elastic neutrino-proton interactions in argon that will allow us to determine the
strangeness contribution to the proton axial form factor and thus provide a value for the strangeness
contribution to the proton spin. This will be the first such measurement in a liquid argon neutrino detector.

\bibliography{INPC2016_Pate}

\begin{thebibliography}{10}

\bibitem{AguilarArevalo:2010cx}
{\sc A.~Aguilar-Arevalo et~al.}, {\em {Measurement of the Neutrino
  Neutral-Current Elastic Differential Cross Section on Mineral Oil at $E_\nu
  \sim 1$ GeV}}, Phys. Rev., D82 (2010), p.~092005.

\bibitem{Ahrens:1987xe}
{\sc L.~A. Ahrens et~al.}, {\em {Measurement of neutrino-proton and
  antineutrino-proton elastic scattering}}, Phys. Rev., D35 (1987), p.~785.

\bibitem{Aidala:2012mv}
{\sc C.~A. Aidala, S.~D. Bass, D.~Hasch, and G.~K. Mallot}, {\em {The Spin
  Structure of the Nucleon}}, Rev. Mod. Phys., 85 (2013), pp.~655--691.

\bibitem{PhysRevC.76.035205}
{\sc J.~Arrington, W.~Melnitchouk, and J.~A. Tjon}, {\em Global analysis of
  proton elastic form factor data with two-photon exchange corrections}, Phys.
  Rev. C, 76 (2007), p.~035205.

\bibitem{PhysRevD.85.054510}
{\sc R.~Babich, R.~C. Brower, M.~A. Clark, G.~T. Fleming, J.~C. Osborn,
  C.~Rebbi, and D.~Schaich}, {\em Exploring strange nucleon form factors on the
  lattice}, Phys. Rev. D, 85 (2012), p.~054510.

\bibitem{Bluemlein2010205}
{\sc J.~Bl{\"u}mlein and H.~B{\"o}ttcher}, {\em {QCD} analysis of polarized
  deep inelastic scattering data}, Nuclear Physics B, 841 (2010), pp.~205 --
  230.

\bibitem{Bodek:2007ym}
{\sc A.~Bodek, S.~Avvakumov, R.~Bradford, and H.~S. Budd}, {\em {Vector and
  Axial Nucleon Form Factors:A Duality Constrained Parameterization}}, Eur.
  Phys. J., C53 (2008), pp.~349--354.

\bibitem{Cabibbo:2003cu}
{\sc N.~Cabibbo, E.~C. Swallow, and R.~Winston}, {\em {Semileptonic hyperon
  decays}}, Ann. Rev. Nucl. Part. Sci., 53 (2003), pp.~39--75.

\bibitem{PhysRevD.80.034030}
{\sc D.~de~Florian, R.~Sassot, M.~Stratmann, and W.~Vogelsang}, {\em Extraction
  of spin-dependent parton densities and their uncertainties}, Phys. Rev. D, 80
  (2009), p.~034030.

\bibitem{Ellis:2008hf}
{\sc J.~R. Ellis, K.~A. Olive, and C.~Savage}, {\em {Hadronic Uncertainties in
  the Elastic Scattering of Supersymmetric Dark Matter}}, Phys. Rev., D77
  (2008), p.~065026.

\bibitem{PhysRevD.86.114510}
{\sc M.~Engelhardt}, {\em Strange quark contributions to nucleon mass and spin
  from lattice {QCD}}, Phys. Rev. D, 86 (2012), p.~114510.

\bibitem{Garvey:1992cg}
{\sc G.~T. Garvey, W.~C. Louis, and D.~H. White}, {\em Determination of proton
  strange form factors from \ensuremath{\nu} \textit{p} elastic scattering},
  Phys. Rev., C48 (1993), pp.~761--765.

\bibitem{Gasenzer:2011hs}
{\sc T.~Gasenzer, O.~Nachtmann, and M.-I. Trappe}, {\em {Metastable states of
  hydrogen: their geometric phases and flux densities}}, Eur. Phys. J., D66
  (2012), p.~113.

\bibitem{Hirai2009106}
{\sc M.~Hirai and S.~Kumano}, {\em Determination of gluon polarization from
  deep inelastic scattering and collider data}, Nuclear Physics B, 813 (2009),
  pp.~106 -- 122.

\bibitem{Karagiorgi:2012cba}
{\sc G.~Karagiorgi}, {\em {MicroBooNE and the Road to Large Liquid Argon
  Neutrino Detectors}}, Phys. Procedia, 37 (2012), pp.~1319--1323.

\bibitem{Leader:2014uua}
{\sc E.~Leader, A.~V. Sidorov, and D.~B. Stamenov}, {\em {New analysis
  concerning the strange quark polarization puzzle}}, Phys. Rev., D91 (2015),
  p.~054017.

\bibitem{Melson:2015spa}
{\sc T.~Melson, H.-T. Janka, R.~Bollig, F.~Hanke, A.~Marek, and B.~Müller},
  {\em {Neutrino-driven Explosion of a 20 Solar-mass Star in Three Dimensions
  Enabled by Strange-quark Contributions to Neutrino-nucleon Scattering}},
  Astrophys. J., 808 (2015), p.~L42.

\bibitem{Nocera:2014gqa}
{\sc E.~R. Nocera, R.~D. Ball, S.~Forte, G.~Ridolfi, and J.~Rojo}, {\em {A
  first unbiased global determination of polarized PDFs and their
  uncertainties}}, Nucl. Phys., B887 (2014), pp.~276--308.

\bibitem{Pate:2012wg}
{\sc S.~Pate, J.~Schaub, and D.~Trujillo}, {\em {Strange Quark Contribution to
  the Nucleon Spin from Electroweak Elastic Scattering Data}}, AIP Conf.Proc.,
  1560 (2013), pp.~503--505.

\bibitem{Pate:2014kea}
{\sc S.~Pate and D.~Trujillo}, {\em {Strangeness Vector and Axial-Vector Form
  Factors of the Nucleon}}, EPJ Web Conf., 66 (2014), p.~06018.

\bibitem{Pate:2003rk}
{\sc S.~F. Pate}, {\em {Determination of the strange form factors of the
  nucleon from $\nu p$, $\bar{\nu}p$, and parity-violating $\vec e p$ elastic
  scattering}}, Phys. Rev. Lett., 92 (2004), p.~082002.

\bibitem{Pate:2008va}
{\sc S.~F. Pate, D.~W. McKee, and V.~Papavassiliou}, {\em {Strange Quark
  Contribution to the Vector and Axial Form Factors of the Nucleon: Combined
  Analysis of G0, HAPPEx, and Brookhaven E734 Data}}, Phys. Rev., C78 (2008),
  p.~015207.

\end{thebibliography}

\end{document}